%% file: main.tex
\documentclass[manuscript,sigconf,10pt]{acmart}

\usepackage{xspace}

\newcommand{\ang}[1]{$#1^\circ$}
\usepackage[english]{babel}
\usepackage{blindtext}
\usepackage{subcaption}
\usepackage{hyperref}
\usepackage{cleveref}
\usepackage{enumitem}
\usepackage{pifont}
\newcommand{\cmark}{\ding{51}}%
\newcommand{\xmark}{\ding{55}}%
\renewcommand\footnotetextcopyrightpermission[1]{} % removes footnote with conference info
\setcopyright{none}
\acmDOI{}

\acmISBN{}

\acmConference[Submitted for review to Mobisys]{}
\acmYear{2026}
\newcommand{\name}{SpyDir\xspace}

\begin{document}

% \title{\name: Accurate \underline{Dir}ection finding of \underline{Spy} Devices}
\title{\name: \underline{Spy} Device Localization Through Accurate \underline{Dir}ection Finding}

\author{Wenhao Chen}
\email{wec035@ucsd.edu}
\affiliation{
  \institution{UC San Diego}
  % \city{College Park}
  \country{USA}
}

\author{Wenyi Morty Zhang}
\email{wez049@ucsd.edu}
\affiliation{
  \institution{UC San Diego}
  % \city{La Jolla}
  \country{USA}
}

\author{Wei Sun}
\email{wei.sun@wright.edu}
\affiliation{
  \institution{Wright State University}
  % \city{}
  \country{USA}
}

\author{Dinesh Bharadia}
\email{dineshb@ucsd.edu}
\affiliation{
  \institution{UC San Diego}
  % \city{La Jolla}
  \country{USA}
}

\author{Roshan Ayyalasomayajula}
\email{roshana@buffalo.edu}
\affiliation{
  \institution{University at Buffalo}
  % \city{}
  \country{USA}
}

\renewcommand{\shortauthors}{Wenhao Chen et al.}

\input{abstract}


%%x
\maketitle

\input{intro/Intro_morty}

\input{design/design_morty}

\input{evaluation/eva_morty}

\input{related/related_morty}
\input{discussion}

\bibliographystyle{ACM-Reference-Format}
\bibliography{ref}
\end{document}

%% file: abstract.tex
\begin{abstract}
 Hidden spy devices (e.g., spy cameras) have become a great privacy threat recently, as these low-cost, low-power, and small form-factor IoT devices can quietly monitor human activities in the indoor environment without generating any side-channel information (e.g., wireless communication traffic and thermal noise). As such, it is difficult to detect and even more challenging to localize them in the rich-scattering indoor environment. To this end, this paper presents the design, implementation, and evaluation of \name, a system that can accurately localize the hidden spy IoT devices by harnessing the electromagnetic emanations automatically and unintentionally emitted from them. Our system design mainly consists of (1) a portable switching antenna array to sniff the spectrum-spread emanations, (2) an emanation enhancement algorithm through non-coherent averaging that can de-correlate the correlated noise effect due to the square-wave emanation structure, and (3) a multipath-resolving algorithm that can exploit the relative channels using a novel optimization-based sparse AoA derivation. Our real-world experimental evaluation across different indoor environments demonstrates an average AoA error of \ang{6.30}, whereas the baseline algorithm yields \ang{21.06}, achieving over a $3.3\times$ improvement in accuracy, and a mean localization error of 19.86cm over baseline algorithms of 206.79cm (MUSIC) and 294.75cm (SpotFi), achieving over a $10.41\times$ and $14.8\times$ improvement in accuracy.
\end{abstract}

% \keywords{Do, Not, Us, This, Code, Put, the, Correct, Terms, for,
  % Your, Paper}

%% file: intro/Intro_morty.tex
\section{Introduction}

The rapid growth of the Internet of Things (IoT) has popularized small, low-power, and low-cost devices. However, the compact and easily deployable nature also makes IoT devices like microcameras easy to misuse for covert monitoring, which creates serious privacy risks. Recent studies show that hidden cameras are increasingly found in hotels, Airbnbs, and private homes to record indoor activities~\cite{airbnb,pcmag}. Detecting and further localizing such covert IoT devices, particularly hidden cameras, is therefore essential for protecting user privacy.

\begin{figure}
    \centering
    \includegraphics[width=0.95\linewidth]{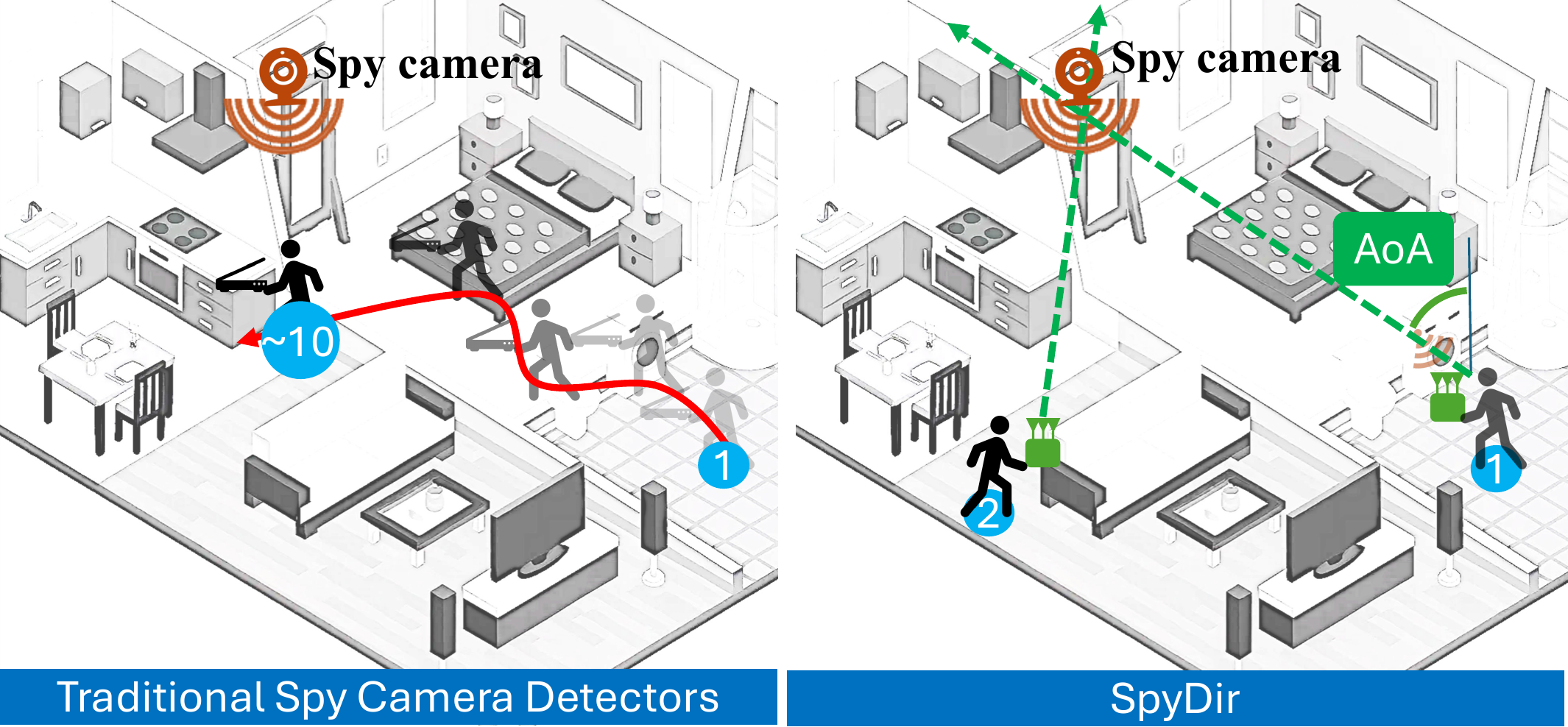}
    \caption{\name (right) can accurately locate a spy camera in just two steps, where it can sniff and derive the emanation direction through AoA estimation and further accurately localize the spy camera. Unlike traditional systems (left), which need you to scan multiple locations to detect and coarsely locate the spy camera based on the camera's thermal noise, wireless communication traffic, or laser-based sensing, etc.}
    \label{fig:system:example}
\end{figure}

%However, detecting the presence of hidden IoT devices in indoor environments through electromagnetic emanation side channel is increasingly feasible, as demonstrated by prior work using dedicated detectors~\cite{ded:detector,anti_detector}, specialized sensors~\cite{he2018active,li2015fast}, traffic analysis~\cite{cheng2018dewicam,singh2021always}, and device emanations~\cite{zhou2023dehirec,sun2025revealing,liu2023camradar}. However, these approaches primarily reveal whether hidden devices exist rather than where they are located. 

\begin{table*}
\centering
\Large
\resizebox{\linewidth}{!}{
\begin{tabular}{|c|c|c|c|c|}
\hline
Detection Model &  \textbf{Protocol-agnostic} & \textbf{Localization-capable} & \textbf{Diverse detection}\\ \hline
Dedicated sensors~\cite{ded:detector} & \cmark & \xmark & \xmark \\ \hline
Traffic-based~\cite{cheng2018dewicam}  &  \xmark & \cmark & \xmark \\ \hline
Sensor-based (optical, laser, and thermal sensors)~\cite{zuniga2022see, he2018active, li2015fast, qian2015recognition} &  \cmark & \xmark & \xmark \\ \hline
Emanation-based~\cite{zhang2024eye} &  \cmark & \xmark & \cmark \\ \hline
\textbf{\textbf{\name}} &  \cmark & \cmark & \cmark \\ \hline
\end{tabular}
}
\caption{System property comparison of \name and prior works.}
\label{tb:comparison}
\end{table*}

As shown in Table~\ref{tb:comparison}, the prior works either lack protocol-agnostic, localization-capable, or diverse detection (i.e., different spy camera detection) properties. For example, the dedicated detectors~\cite{ded:detector,anti_detector} or specialized sensors~\cite{li2015fast} cannot detect diverse hidden spy IoT devices.  Moreover, it is difficult to detect these low-cost, low-power, and small form-factor hidden spy IoT devices, as they may not generate any side-channel information, such as wireless communication traffic and very little thermal noise, which can disable most traffic-based~\cite{cheng2018dewicam,singh2021always}, thermal imaging-based detections~\cite{zuniga2022see}, and optical/laser sensor-based detections~\cite{he2018active}. Recent works exploit the electromagnetic emanations~\cite{zhou2023dehirec,sun2025revealing,liu2023camradar} that are automatically and unintentionally emitted from the hidden cameras for detection. For example, RFScan~\cite{sun2025revealing} localizes hidden devices by tracing their emanation harmonics with a directional antenna, yet its accuracy degrades significantly in multipath-rich indoor environments. Similarly, ESauron~\cite{zhang2024eye} depends on the RSSI of electromagnetic radiation and suffers from parallel limitations under complex propagation conditions. CamFirm~\cite{liao2025camfirm} induces electromagnetic responses by actively modulating ambient brightness during camera mode switching for localization. However, this approach only functions reliably in relatively dim lighting, which restricts its practical applicability in typical indoor scenarios. Therefore, localizing hidden cameras from their unintentional emanations in complex real-world environments remains an open challenge, due to strong multipath effects and the inherently low signal-to-noise ratio of these emanations. Considering these constraints in realistic indoor scenarios, we identify and address three key challenges that hinder accurate localization:

\noindent\textit{(1) A portable and low-cost array for emanation eavesdropping.} Hidden cameras emit amplitude-modulated clock signals whose emanation spikes appear across a wide and discontinuous sub-GHz spectrum. Accurately localizing these devices requires capturing these distributed spikes while retaining the spatial resolution needed for Angle-of-Arrival (AoA) estimation. A conventional solution is to use a bulky multi-antenna array with multiple RF chains, but such setups are expensive and impractical. To make localization feasible with low-cost and readily available hardware, \textit{we design a handheld switching-antenna array PCB that uses only a synchronized dual-channel RF front-end (e.g., a standard USRP).} This design remains compact and inexpensive while still achieving the high-speed switching necessary to derive the relative channels used for emanation-based AoA estimation and localization.

\noindent\textit{(2) Emanation enhancement, noise decoupling, and interference detection.} Unintentional clock emanations from hidden cameras are inherently very weak, making their spectral spikes difficult to detect and even harder to use for localization. Non-coherent averaging would ideally boost the signal strength. However, in a switching-antenna architecture, we observed in practice that the noise introduced at different antenna positions is correlated, preventing effective signal enhancement. To address this, we adapt non-coherent averaging by introducing controlled delays that leverage the periodic square-wave structure of the emanations. \textit{These delays drive the noise components out of phase and reduce their correlation, while the structured emanation remains consistently reinforced.} As a result, the switching noise is mitigated, and the effective emanation energy is boosted.

\noindent\textit{(3) AoA estimation with sparse and joint optimization.} Indoor environments introduce strong multipath that significantly distorts the phase of the received emanations, making localization difficult in cluttered spaces. Although a switched antenna array can, in principle, estimate AoA through triangulation, unintentional emanations do not contain pilots, so conventional CSI (Channel State Information)–based approaches, such as MUSIC~\cite{schmidt1986multiple}, cannot be applied. Inspired by SpotFi~\cite{kotaru2015spotfi}, we extract relative channels by comparing each switched antenna element with a fixed reference antenna. However, the resulting smoothed relative CSI matrix has limited rank, which prevents effective separation of multiple paths. To address this limitation, \textit{we propose a sparse optimization algorithm that recovers multiple AoAs more reliably than SpotFi-style smoothing.} Specifically, we construct multiple overlapping subspaces to artificially increase the rank and improve multipath resolvability, which can partially compensate for the limitation introduced by the relative-channel structure. 

In this paper, we propose \name, a system that accurately localizes hidden cameras by estimating the AoA of their automatic and unintentional emanations using a portable switching-antenna array, as shown in Fig.~\ref{fig:system:example}. \name monitors the spectrum-spread electromagnetic emanations from \textit{diverse spy cameras w/ or w/o wireless communication architectures}, performs \textit{protocol-agnostic} spectrum analysis, and remains robust under strong indoor multipath, enabling accurate \textit{localization} through a novel noise decorrelation approach and an innovative sparse optimization algorithm. As such, our \name achieves the protocol-agnostic, localization-capable, and diverse devices sensing properties as shown in Table~\ref{tb:comparison}.

We implement \name using software-defined radios (USRPs) and a custom high-speed switching-antenna PCB, and evaluate it across diverse indoor environments and multiple representative hidden-camera devices. Our evaluation spans multiple hidden-camera devices, diverse indoor environments, varying distances and frequencies, and different packet volumes, together with analyses of algorithmic variants and parameter choices. Overall, \name achieves an average AoA error of \ang{6.30}, offering over a $3.3\times$ improvement relative to MUSIC. When deployed for full 2D localization, \name attains an average position error of 19.86 cm, whereas MUSIC and SpotFi exceed 200 cm and 290 cm, respectively—corresponding to 10×–15× localization accuracy improvement in real indoor environments. Our contributions are listed as follows: 
\begin{itemize}[leftmargin=*]
\item To the best of our knowledge, we present the first system that accurately localizes hidden spy IoT devices using unintentional emanations in multipath-rich indoor environments.
\item We design a compact wideband switched-antenna array and signal-enhancement pipeline that boosts extremely weak emanations by combining non-coherent averaging with a novel noise decorrelation approach.
\item We develop a sparse optimization algorithm that derives reliable AoA estimates from relative CSI obtained through antenna switching, which is not possible with the state-of-the-art AoA estimation algorithms.
\item We validate \name across diverse indoor settings, device types, and operating conditions, demonstrating its practicality and significantly improved AoA accuracy compared to existing baselines. We also plan to open-source our implementation and hardware design to facilitate reproducibility and future research.
\end{itemize}

%% file: design/design_morty.tex
\input{design/background_morty}

\input{design/system_overview_morty}

\input{design/system_design_morty}

%% file: design/background_morty.tex
\section{Background}\label{sec:background}

We begin by briefly describing how leakage signals behave and how existing state-of-the-art detection algorithms identify spy devices. These primers clarify the capabilities and limitations of current systems, which focus on detecting such devices but not locating them. We then present in~\cref{sec:design} how \name enables practical localization of spy devices in real-world deployments.

\subsection{Understanding Leakage Signals}\label{sec:back-leakage}

\textbf{Physical principle.}
Hidden spy IoT devices, such as hidden spy cameras, contain CPUs and memory modules driven by internal clock signals. These clock signals, when coupled with computational activities like memory access, produce amplitude-modulated clock signals commonly referred to as electromagnetic emanations or emanations. Although unintentional, these emanations automatically leak through circuit traces, interconnects, and communication lines on the device, which effectively behave as small radio-frequency antennas. As a result, hidden cameras emit detectable RF leakage signals without any modification to their hardware or firmware.

\begin{figure}[t]
    \centering
    \includegraphics[width=\linewidth]{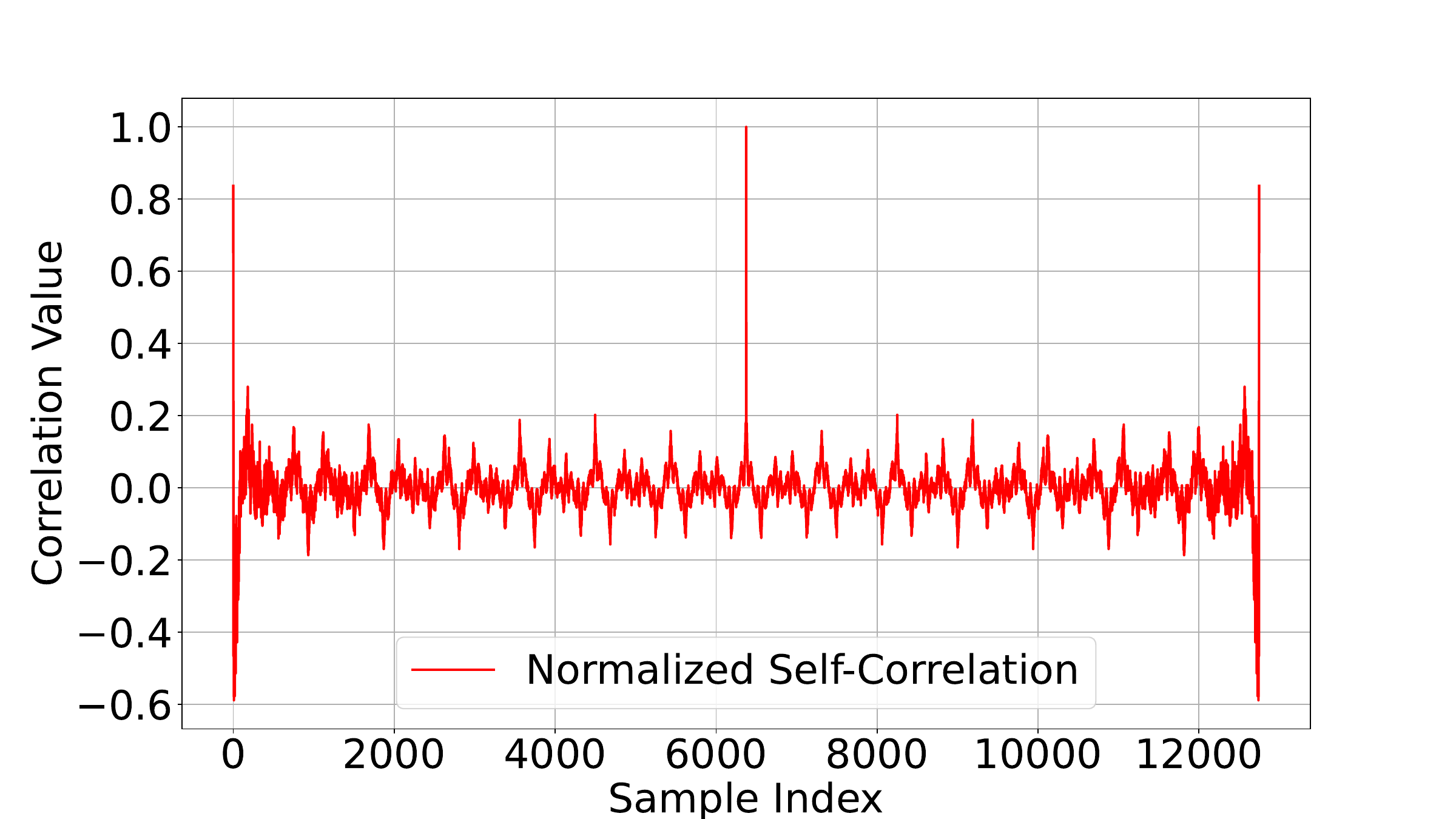}
    \caption{Autocorrelation of an emanation signal showing clear harmonics of the device’s internal clock.}
    \label{fig:back-detect}
\end{figure}

\noindent\textbf{Emanation characteristics.}
In the time domain, the emanations appear as square-wave patterns because the internal clock is amplitude-modulated by the device’s computational activity. In the frequency domain, these signals spread across a wide sub-GHz band and produce multiple harmonics/spikes, including the fundamental frequency and its higher-order multiples. Due to this spectral spreading, the harmonic components are weak, and their strength decreases with frequency. The spacing between adjacent harmonics reflects the underlying clock frequency, which can be identified through the autocorrelation of the received IQ samples. As shown in Fig.~\ref{fig:back-detect}, the autocorrelation exhibits periodic peaks corresponding to the clock frequency, enabling both emanation detection and subsequent localization.

\subsection{Limitations of Existing Leakage Detection}\label{sec:back-detect}

To detect emanations, the goal is to identify the amplitude-modulated clock signals, which appear as square waves in the time domain and as harmonics or spikes in the frequency domain. Previous work~\cite{shen2021earfisher, sun2025revealing, zhou2023dehirec, liu2023camradar} typically detects these harmonics by first applying a Fourier transform and then locating the spectral spikes corresponding to the device’s clock frequency. Accurate clock-frequency estimation requires identifying all harmonics across a wide frequency band. However, higher-order harmonics are often much weaker, and emanations in general have inherently low power. As a result, existing systems often rely on directional antennas, power amplification, or extensive spectral averaging to boost these weak spikes before detection. Once the emanation spikes are detected and the clock frequency is derived, the presence of concealed spy devices within the scene can be confirmed. 

However, these approaches focus exclusively on detection rather than localization, and they depend on hardware or averaging techniques that do not generalize well to practical AoA estimation under weak unintentional signals, strong multi-path, and low SNR in real indoor environments. Addressing these limitations requires a fundamentally different design. In this paper, we tackle both the weak-signal problem and the challenge of estimating the blind source signal AoA for localization, and develop \name, which is described in the next section.

%% file: design/system_overview_morty.tex
\section{System Overview}

\begin{figure}[t]
    \centering
    \includegraphics[width=\linewidth]{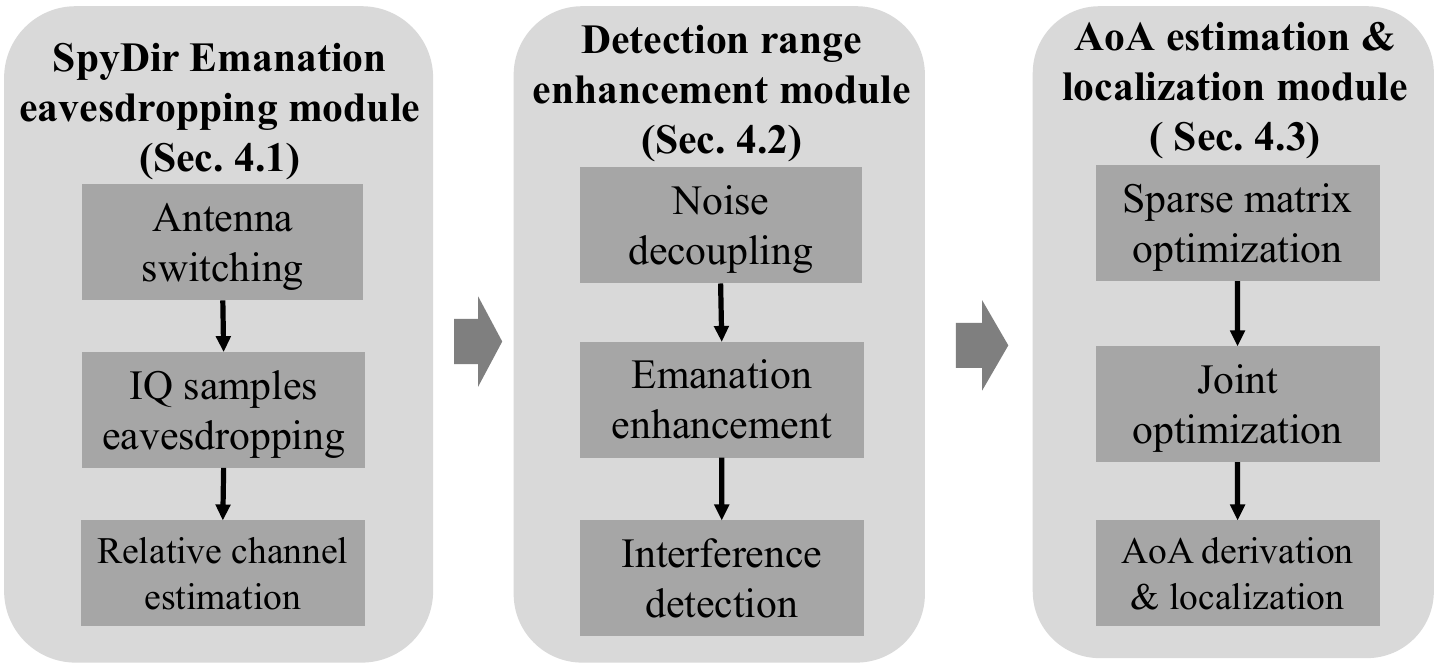}
    \caption{The workflow of \name consists of three modules: the switched-antenna array module, the detection enhancement module, and the AoA estimation and localization module.}
    \label{fig:workflow}
\end{figure}

Fig.~\ref{fig:workflow} shows the workflow of \name, which consists of three modules: 1) the \name emanation eavesdropping module, 2) the detection enhancement module, and 3) the AoA estimation and localization module. During localization, the switched-antenna array passively captures emanation signals from hidden spy devices in the indoor environment, and our algorithms process these signals to determine the device’s exact location. We summarize the function of each module below:

\begin{itemize}[leftmargin=*]

\item \textbf{\name emanation eavesdropping module~(\S\ref{sec:array_design}).}
To analyze the emanations emitted from the spy cameras, we leverage a compact switched-antenna array that uses only two RF ports, allowing direct use with commodity USRPs while still enabling estimation of the relative channels between the reference antenna and each antenna element. As such, our \name is portable for ubiquitous spy camera localization in any indoor environment. 

\item \textbf{Detection enhancement module~(\S\ref{sec:time_offset_design}).}  
Using the estimated channels, we enhance the weak emanation signals through non-coherent averaging and a delay-based noise decorrelation technique, extending the effective detection range to up to $3.6$~m while maintaining an AoA error below $20^{\circ}$. Moreover, we also detect the diverse ambient emanation interferences and further eliminate them for accurate relative channel estimation.

\item \textbf{AoA estimation and localization module~(\S\ref{sec:AOA_design}).}  
With enhanced relative channels, we apply a sparse matrix optimization algorithm, including a sparse optimization and a joint optimization, to accurately estimate the AoA of the emanations and localize the hidden camera.

\end{itemize}

%% file: design/system_design_morty.tex
\section{System Design}\label{sec:design}

% \name's system design revolves around the need for not just identifying but also in locating these Spy devices in any given environment at a reasonable form factor. In designing \name, we made contributions towards a switched antenna relative channel measurement device that we first discuss in ~\cref{sec:array_design}. This array design help us get channel measurements that enable in angle-of-arrival measurements for the incoming signal. We then tackle the below noise floor signals for most of thee emanations for long range ($~1-2$~m.) detections. To this end we design phase-offset based coherent averaging algorithm described in ~\cref{sec:time_offset_design} that enabled high-SNR for emanation signals extending the detection and AoA estimation range from $<1$~m to $2.2$~m. Finally, we design a sparse-matrix based optimization to estimate accurate AoA in ~\cref{sec:AOA_design} that provides accurate direction finding for the Spy device.

% \begin{figure}[t]
%     \centering
%     \includegraphics[width=0.9\linewidth]{figures/spydir_switch_antena-crop.pdf}
%     \caption{Schematic of the switched-antenna array and switching cycles enabling multi-antenna emanation capture for AoA estimation.}
%     \label{fig:design-switched-array}
% \end{figure}

\subsection{\name Emanation Eavesdropping}\label{sec:array_design}

As described in~\S\ref{sec:back-detect}, detecting emanations simply involves identifying the repetitive harmonics in the frequency domain that arise from the device’s square-wave clock. In contrast, estimating the angle of arrival (AoA) is significantly more challenging, as it requires coherent phase information across antennas and samples. This corresponds to a blind-source estimation problem, where the transmitted waveform is unknown~\cite{luo2018comprehensive,yu2014blind}. To address this, we want to leverage a portable mobile device that can eavesdrop on the emanations from the spy camera.   

\begin{figure}
    \centering
    \includegraphics[width=0.9\linewidth]{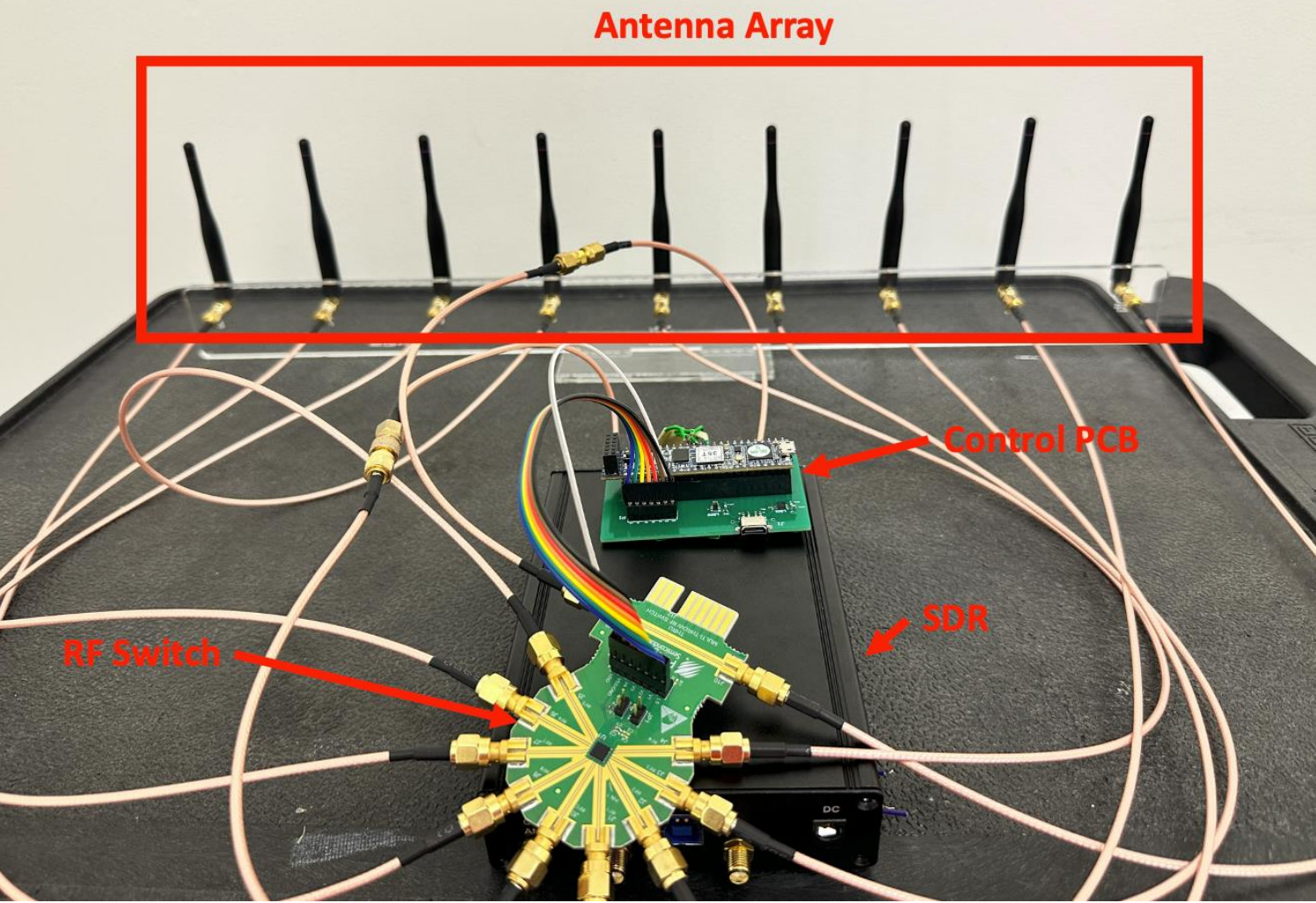}
    \caption{\name consists of the RF switch, SDR, control PCB, and an antenna array.}
    \label{fig:design-switched-array}
\end{figure}

\noindent \textbf{Primer on Switched-Antenna Array.} To do so, we leverage a switched-reference antenna array in which one antenna serves as a fixed reference while the remaining antennas are connected to a high-speed RF switch, as shown in Fig.~\ref{fig:design-switched-array}. Then, we can switch to different antennas while maintaining only two RF ports, which enables us to use a two-port software-defined radio (SDR) for emanation eavesdropping. The wireless signals received across the switched antennas can be used to emulate an antenna-array-based signal processing. Since it only requires two RF chains for signal processing, we can use a simple and portable SDR (e.g., USRP B210) for emanation signal processing and emanation source localization.

\noindent \textbf{Leverage Switched-Antenna Array for Emanation Eavesdropping.} Our implementation uses both RX ports of a USRP B210 operating in synchronized mode: one RF port is directly connected to the reference antenna, and the other is connected to the RF switch, which sequentially selects among the array elements. This design requires only one inexpensive USRP with two RX ports, instead of multi-channel or multi-USRP synchronized systems typically required for coherent arrays. The internal 40 kHz clock of the B210 is tapped and used to generate synchronized control signals for the switch, ensuring precise alignment between the USRP sampling and the antenna-switching cycle.

Unfortunately, leakage signals rarely satisfy these conditions in our real-world observations. \textit{In practice, the SNR is low, especially as the distance to the spy device increases, and the noise components across the reference and switched antennas are not independent.}  To address this, \name explicitly models the noise coupling and introduces a  phase-offset–based relative channel estimator, which decorrelates the noise  components and significantly extends the effective leakage-signal detection  range (by more than \( 100\% \)), as discussed in the next section.
% This high noise correlation prevents simple temporal averaging (i.e., estimating \( \mathbb{E}[\cdot] \)) from producing accurate narrowband channel estimates for AoA estimation.

\begin{figure}[t]
    \centering
    \includegraphics[width=0.95\linewidth]{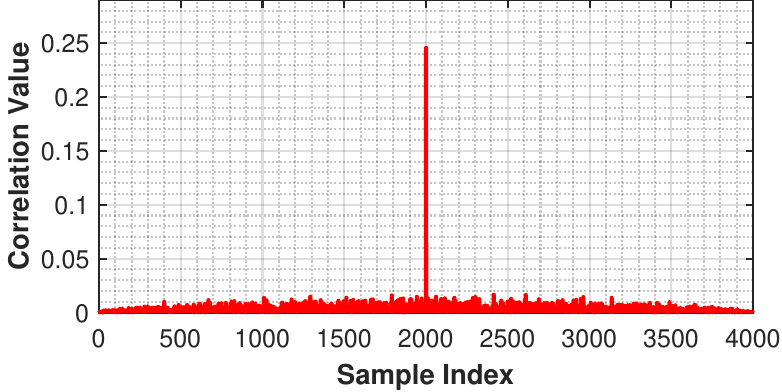}
    \caption{Noise cross-correlation between the switched antenna and the reference antenna, measured in an anechoic chamber. With no external signals present, the received samples contain only hardware and environmental noise, revealing strong noise correlation across antennas.}
    \label{fig:high-noise-corr}
\end{figure}

% \begin{equation}
% \begin{aligned}
% &S(\tau)= \mathbb{E}[s(t)s^*(t-\tau)],\\
% &C_i(\tau)=\mathbb{E}[s(t)n_i^*(t-\tau)],\\
% &N_{i,k}(\tau)=\mathbb{E}[n_i(t)n_k^*(t-\tau)],\\
% \end{aligned}
% \end{equation}

% When the SNR is high and the noise is independent, this estimator can provide an accurate channel estimation. 

\subsection{Enhancing Detection Range}\label{sec:time_offset_design}

\subsubsection{Relative Channel Estimation Paradigm}

In switched-antenna array systems, the device does not measure the CSI directly. 
Instead, it obtains the relative channel by computing the phase and amplitude shift 
between the reference antenna and each switched antenna. Let $r_{\text{ref}}(t)$ denote the 
received signal at the reference antenna. For the $i$th antenna in an $N_{\text{ant}}$-element 
switched array ($i \in \{1,2,\dots,N_{\text{ant}}\}$), let $r_{i,\text{switch}}(t)$ be its 
received signal. A widely used estimator for extracting the relative narrowband channel is:

\begin{equation} \label{eq:1}
h_i^{est} 
=  \frac{\mathbb{E}[r_{i,\text{switch}}(t) r_{\text{ref}}^*(t)]}{\mathbb{E}[|r_{\text{ref}}(t)|^2]}.
\end{equation}

where \( h_i^{est} \) represents the estimated relative channel of the \( i \)th antenna.  \( \mathbb{E}[\cdot] \) is the expectation operator over time. For narrow-band signals the received samples can be further expressed in terms of signal and noise as follows:
\begin{equation}
r_{i,\text{switch}}(t) = h_i s(t) + n_i(t), \quad r_{\text{ref}}(t) = s(t) + n_{\text{ref}}(t)
\end{equation}

where $h_i$, $s(t)$, and $n_i(t)$ denote the true relative channel, the target signal, and the noise at the $i$th switch antenna, respectively. Substituting the signal and noise components into Eq.~\eqref{eq:1}, we obtain:

\begin{equation}\label{eq:3}
\begin{aligned}
h_i^{est}
&=  \frac{\mathbb{E} [(h_i s(t) + n_i(t)) (s^*(t) + n_{\text{ref}}^*(t))]}{\mathbb{E} [(s(t) + n_{\text{ref}}(t)) (s^*(t) + n_{\text{ref}}^*(t))]} \\
&= \frac{h_i S(0)+h_i C_{ref}(0)+C_i^*(0)+N_{i,ref}(0)}{S(0)+ C_{ref}(0)+C_{ref}^*(0)+N_{ref,ref}(0)} 
\end{aligned}
\end{equation}
where $S(\tau) = \mathbb{E}[s(t)s^*(t-\tau)]$ is the autocorrelation matrix for the signal ($s(t)$), $C_i(\tau)=\mathbb{E}[s(t)n_i^*(t-\tau)]$ is the cross-correlation between the signal and noise measured at the $i^{th}$ antenna ($n_i(t)$), and $N_{i,k}(\tau)=\mathbb{E}[n_i(t)n_k^*(t-\tau)]$ is the cross correlation between the noise components of the measurements at $i^{th}$ and $k^{th}$ antenna. Now looking at Eq~\ref{eq:3} we can see that only when the SNR is high and the noise across reference and switch antenna are independent, $h_i^{est}$ reduces to $h_i$, which is the true relative channel we wanted to measure\footnote{In this case, \( C(0) \) and \( N(0) \) can be ignored, with \( S(0) \) being the dominant term.}.

\subsubsection{Noise Coupling in Low SNR Scenarios} 

Hhowever, as discussed earlier, achieving high SNR is difficult in our setting because clock-leakage signals inherently carry very little energy. In addition, we observe that unavoidable factors such as antenna coupling and shared hardware paths cause the noise received at different antennas to exhibit strong correlation, as shown in Fig.~\ref{fig:high-noise-corr}. Although the clock signal approximates a square wave in the time domain (yielding \( C(\tau) \approx \mathbb{E}[N(t - \tau)] \to 0 \), allowing \( C(0) \) to be ignored), the strong cross-antenna noise correlation remains problematic. This correlation produces a biased estimator for $h_i^{est}$, as derived in Eq.~\ref{eq:biased estimator}:
\begin{equation} \label{eq:biased estimator}
\begin{aligned}
h_i^{est}
&= \frac{h_i S(0)+N_{i,ref}(0)}{S(0)+N_{ref,ref}(0)} \\
&=h_i\frac{ S(0)}{S(0)+N_{ref,ref}(0)}+\frac{N_{i,ref}(0)}{S(0)+N_{ref,ref}(0)}
\end{aligned}
\end{equation}
The resulting bias increases as the SNR decreases, making accurate channel estimation more difficult in low-SNR conditions.

% From the above equation, we can see that the bias 

\subsubsection{SNR Enhancement}\label{subsec:time_offset_design}

The most common method for improving SNR is coherent averaging~\cite{yousef2003robust}, which requires collecting a large number of phase-aligned packets and averaging them. However, synchronizing the initial phase of a low-energy and unknown waveform across packets is extremely difficult. The limited coherence time of the channel further restricts the applicability of this technique.

In this section, we propose a time-offset estimator that significantly enhances SNR without increasing the number of received samples. We observe that the clock-leakage signal has a nearly constant period, which leads to strong periodic autocorrelation. In contrast, the cross-correlation of noise closely resembles a unit impulse, $\delta(t)$. By exploiting these properties, we introduce a carefully chosen time offset into the standard estimator:
\begin{equation}  \label{eq:offset_estimator}
\begin{aligned}
h_{i,\tau}^{\text{offset}} 
&=  \frac{\mathbb{E}[r_{i,\text{switch}}(t) r_{\text{ref}}^*(t-\tau)]}{\mathbb{E}[r_{\text{ref}}(t) r_{\text{ref}}^*(t-\tau)]]}\\
&= \frac{h_i S(\tau)+h_i C_{ref}(\tau)+C_i^*(\tau)+N_{i,ref}(\tau)}{S(\tau)+ C_{ref}(\tau)+C_{ref}^*(\tau)+N_{ref,ref}(\tau)} 
\end{aligned}
\end{equation}

As discussed earlier, because the clock-leakage signal closely approximates a square wave, the term \( C(\tau) \) can be treated as negligible. Substituting this approximation into the estimator, we can get:
\begin{equation}
h_{i,\tau}^{\text{offset}}=h_i\frac{ S(\tau)}{S(\tau)+N_{ref,ref}(\tau)}+\frac{N_{i,ref)}(\tau)}{S(\tau)+N_{ref,ref}(\tau)}
\end{equation}

In this case, the effective SNR becomes \( \frac{S(\tau)}{N(\tau)} \). By selecting an appropriate value of \( \tau \), we can theoretically increase the SNR indefinitely. As discussed before, the autocorrelation and cross-correlation of noise approximate a unit-impulse function, $\delta(t)$, while the target signal exhibits strong periodic autocorrelation properties. In fact, choosing \( \tau \) as an integer multiple of the target signal's period often yields good results. Consequently, selecting \( \tau \) as an integer multiple of the signal period often yields favorable results.

\subsubsection{Interference Detection}\label{subsec:interference}

Clock leakage occurs at specific frequencies that often coincide with emissions from other electronic devices. As a result, interference is unavoidable in realistic indoor environments. In this section, we analyze how interference affects our estimator and describe how \name detects the presence of interference.

Let the target leakage signal be \( s(t) \) and an interfering leakage signal be \( d(t) \). The signal received at the reference antenna is:
\[
r_{\text{ref}}(t) = s(t) + d(t),
\]
and the signal received at the \( i \)-th switched antenna is:
\[
r_i(t) = h_i s(t) + \alpha_i d(t),
\]
where \( h_i \) and \( \alpha_i \) denote the channels of the target and interference signals, respectively. For clarity, we omit noise terms and focus on how interference interacts with our SNR-enhancement estimator. The estimated channel becomes:
\begin{equation} \label{eq:inteference}
\begin{aligned}
h_{i,\tau}^{\text{offset}}
&= h_i \frac{S(\tau) + C(\tau)}
         {S(\tau) + C(\tau) + C^{*}(-\tau) + D(\tau)} \\
&\quad + \alpha_i \frac{C^{*}(-\tau) + D(\tau)}
         {S(\tau) + C(\tau) + C^{*}(-\tau) + D(\tau)}
\end{aligned}
\end{equation}

% where \( C(\tau) = \mathbb{E}[s(t)d^{*}(t-\tau)] \) is the cross-correlation 
% between the target and interference, and 
% \( D(\tau) = \mathbb{E}[d(t)d^{*}(t-\tau)] \) is the interference autocorrelation.

where {\( C(\tau) = \mathbb{E}[s(t)d^*(t-\tau)] \)}. As shown in the equation above, interference simply adds additional channel components, each scaled by a normalized weight. In other words, interference only introduces extra AoA paths rather than corrupting or destroying the true target direction. Based on this observation, \name detects the presence of interference by also computing the inverse-offset estimate $h_{i,\tau}^{\text{inverse}}$ from Eq.~\ref{eq:offset_estimator_inverse} and comparing it with the expression in Eq.~\ref{eq:inteference}:
\begin{equation}  \label{eq:offset_estimator_inverse}
\begin{aligned}
h_{i,\tau}^{\text{inverse}}
&=  \frac{\mathbb{E}[r_{i,\text{switch}}(t-\tau) r_{\text{ref}}^*(t)]}{\mathbb{E}[r_{\text{ref}}(t-\tau) r_{\text{ref}}^*(t)]}\\
&= h_i \frac{S^*(\tau)+C(-\tau)}{S^*(\tau)+ C^*(\tau)+C(-\tau)+D^*(\tau)}\\
&+\alpha_i \frac{C^*(\tau)+D^*(\tau)}{S^*(\tau)+ C^*(\tau)+C(-\tau)+D^*(\tau)}
\end{aligned}
\end{equation}

In the absence of interference, both Eq.~\ref{eq:inteference} and Eq.~\ref{eq:offset_estimator_inverse} reduce to $h_{i,\tau}^{\text{offset}} = h_{i,\tau}^{\text{inverse}} = h_i$. Therefore, any deviation between $h_{i,\tau}^{\text{offset}}$ and $h_{i,\tau}^{\text{inverse}}$ indicates the presence of interference. In such cases, we can re-estimate the channel using samples from a different time segment. This mechanism makes \name robust to environmental interference and enables it to detect and localize devices whose leakage signals may otherwise be masked by stronger emissions.

% Thus, our algorithm to estimate the relative channel using additional phase offset can also aid us in identifying any interference. That is In addition, we propose a simple method to detect the presence of interference. Apart from the channel estimation in eq(\ref{eq:inteference}), we also calculate:

% if there is no interference, those two value should be equal. 

\subsection{Estimate Accurate AoA}\label{sec:AOA_design}

Using the proposed estimator, the recovered channel reflects the ratio between the true channels of the switched antenna and the reference antenna. In the presence of multipath, the relative channel observed at the \( i \)-th switched antenna can be expressed as:
\begin{equation}\label{eq:relative_channel}
h_{i} = \frac{\sum_{k=1}^{m} w_k e^{-j 2\pi isin(\theta_k)d/\lambda}}{\sum_{k=1}^{m} w_k }
\end{equation}
where \( m \) denotes the number of propagation paths, \( w_k \) is the complex gain of the \( k \)-th path at the reference antenna, \( \theta_k \) is its angle of arrival (AoA), \( d \) is the antenna spacing, and \( \lambda \) is the signal wavelength. By switching across all \(N\) antennas, we obtain the relative channel vector:
\begin{equation}\label{eq:channel_vector}
\frac{1}{\sum_{k=1}^{m} w_k}
\Big[
\sum_{k=1}^{m} w_k,
\sum_{k=1}^{m} w_k \phi_k,
\sum_{k=1}^{m} w_k \phi_k^2,
\dots,
\sum_{k=1}^{m} w_k \phi_k^{N}
\Big]
\end{equation}
where \(\phi_k=e^{-j 2\pi sin(\theta_k)d/\lambda}\) is the per-antenna phase shift of the \( k \)-th path induced by the array geometry.

\subsubsection{Limitations of SpotFi/MUSIC-Based AoA Estimation}\label{subsec:Spitfi_design}

Directly applying the relative channel vector in Eq.~\ref{eq:channel_vector} to the MUSIC algorithm produces a signal subspace with \(\text{rank} = 1\). This occurs because all antennas observe the same set of multipath components up to a common normalization term. As a result, multipath cannot be effectively separated, and the AoA estimates degrade severely. 

To mitigate this, we consider a SpotFi-inspired approach that constructs multiple overlapping subspaces to artificially increase rank and improve multipath resolvability. This technique trades off the number of usable antennas in order to generate a set of short, full-rank channel segments.

For instance, in a system with 9 antennas (8 switched elements and 1 reference), we can select every 5 consecutive channel values to form a subspace, yielding a \( 5 \times 5 \) covariance matrix with full rank. This improves the ability to resolve multiple propagation paths and partially compensates for the limitations imposed by the relative-channel structure.

\subsubsection{Sparse Matrix Optimization} \label{subsec:optimization_design}

\begin{figure}[t]
    \centering
    \includegraphics[width=0.9\linewidth]{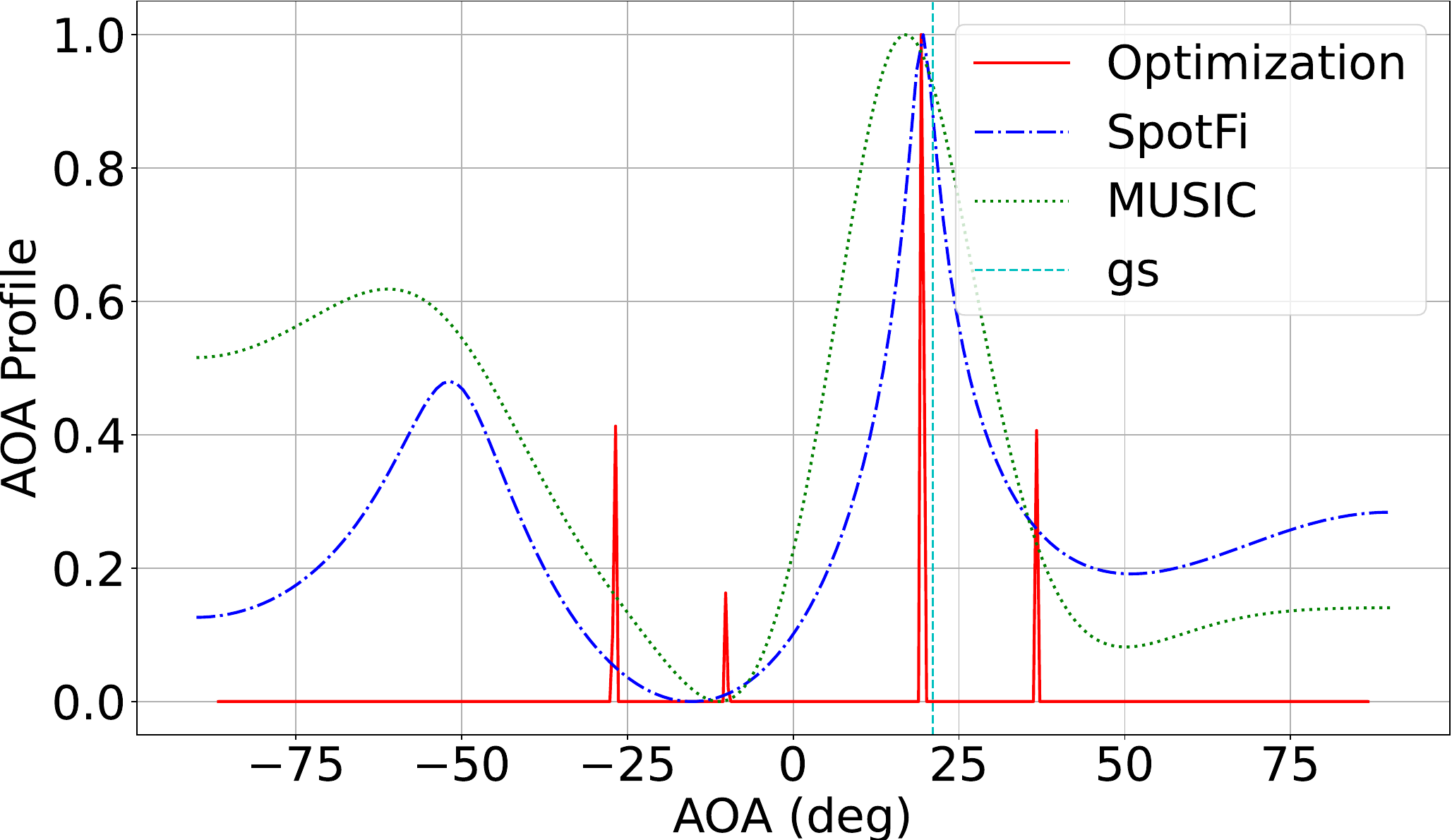}
    \caption{\name’s AoA estimator (labeled “Optimization”) provides significantly higher angular resolution than existing super-resolution methods such as MUSIC~\cite{schmidt1986multiple} and SpotFi~\cite{kotaru2015spotfi}. “gs” denotes the ground-truth direction.}
    \label{fig:design-aoa-res}
\end{figure}

To overcome the low-rank limitations of the estimated channel matrix, which fundamentally restrict the performance of super-resolution algorithms such as MUSIC and SpotFi, we formulate a sparse representation of the AoA profile and solve it using a sparse-matrix optimization procedure. This enables high-resolution AoA estimation, as illustrated in Fig.~\ref{fig:design-aoa-res}.

\smallskip
\noindent\textbf{IFFT and Window Effects.} A classical alternative for obtaining an AoA profile is to apply an inverse fast Fourier transform (IFFT), which maps the antenna domain into the spatial (AoA) domain~\cite{ayyalasomayajula2020deep}. Applying a large-size IFFT to the channel vector in Eq.~\ref{eq:channel_vector} yields:
\begin{equation}
P(\Psi)
= \sum_{k=1}^{m}
\left(
\frac{w_k}{\sum_{k=1}^m w_k}
\delta\left(\Psi - \frac{d \sin(\theta_k)}{\lambda}\right)
\right)*w_{\text{sinc}}(\Psi)
\end{equation}
where $*$ denotes convolution and \(w_{\text{sinc}}(\Psi)\) is the sinc-shaped window function imposed by the finite number of antennas. The width of this window directly limits the resolution of any IFFT-based AoA estimate.

Since the IFFT result is discrete, the convolution can be equivalently rewritten in matrix form as \(\vec P=\vec W\vec a\) where \(\vec P\) is a \(D\times 1\) vector (IFFT result with FFT size \(D\)), \(\vec W\) is a \(D\times D\) window matrix whose columns are shifted versions of \(w_{sinc}(\Psi)\) and \(\vec a\) is a 
\(D\times 1\) sparse vector encoding the underlying AoA components.

A natural approach for estimating \(\vec a\) is to solve the least-squares problem:
% \begin{equation} \label{eq:simple_optmization}
%     \min_{\vec a}  |\vec P-\vec W\vec a|_2
% \end{equation}
\begin{equation}\label{eq:simple_optmization}
\min_{\vec{a}} \lVert \vec{P} - \vec{W}\vec{a} \rVert_2
\end{equation}
However, because the rank of the window matrix \(\vec W\) is fundamentally limited by the number of antennas, the optimization in Eq.~\ref{eq:simple_optmization} becomes undetermined.

\noindent\textbf{Sparse Optimization.} To resolve the undetermined system caused by the limited rank of \(\vec W\), we incorporate structural constraints on the AoA profile. From the IFFT formulation, the true AoA representation can be written as:
\[
\vec{a} = \sum_{k=1}^{m} \frac{w_k}{\sum_{k=1}^{m} w_k} \delta \left( \Psi - \frac{d \sin(\theta_k)}{\lambda} \right)
\]
which reveals several important properties. First, \(\vec{a}\) is sparse: it contains only \(m\) nonzero entries corresponding to the \(m\) propagation paths. Second, the nonzero entries sum to one due to the normalization term. Third, the support of \(\vec{a}\) is restricted to the physically valid region:
\[
|\Psi| = \left|\frac{d\sin(\theta)}{\lambda}\right| \le \frac{d}{\lambda},
\]
because \(|\sin(\theta)| \le 1\). Thus, when the antenna spacing satisfies \( d < \frac{\lambda}{2} \), all entries of \(\vec{a}\) outside the interval \([ -d/\lambda ,\, d/\lambda ]\) must be zero, further reducing the feasible search space.

% This implies that \(\vec{a}\) is a sparse vector, containing only \(m\) non-zero values. Moreover, we can observe that in this formulation, all non-zero values in \(\vec{a}\) should sum up to 1. Additionally, it's easy to note that the non-zero values occur at \(\Psi = \frac{d \sin(\theta_k)}{\lambda}\). Considering that \(-1 \leq \sin(\theta) \leq 1 \) for all \(\theta\), it follows that \(\vec{a}(\Psi) = 0\) for all \(|\Psi| > \frac{d}{\lambda}\). This helps further narrow down the search space when the antenna spacing \(d\) is smaller than \(\frac{\lambda}{2}\).

% Based on these observations, we can now add the constraints to the optimization problem. Where the problem of angle-of-arrival (AoA) estimation can be formulated as:

Leveraging these observations, we impose sparsity and support constraints on \(\vec{a}\) and obtain the following constrained optimization formulation for AoA estimation:
\begin{equation}\label{eq:optimization}
\begin{split}
    \underset{\vec{a}}{\text{min}} & \quad \|\vec{P} - \vec{W} \vec{a}\|_2 + \beta \|\vec{a}\|_1 \\
    \text{s.t.} & \quad \sum \vec{a} = 1 \\
    & \quad \vec{a}(\Psi) = 0, \quad \forall |\Psi| > \frac{d}{\lambda} 
    \end{split}
\end{equation}
where we introduce the \(\ell_1\)-norm of \(\vec{a}\), \(\|\vec{a}\|_1\), to approximate the \(\ell_0\)-norm and ensure a sparse solution, which constraints further tighten the solution. Thus, using this sparse-matrix representation and the constrained optimization formulated in Eq.~\ref{eq:optimization}, we enhance the AoA estimation resolution even for the low-ranked leakage signal measurements across \name's antenna array. Thus enabling accurate direction finding of the Spy Device in any given indoor environment, not only to detect but to identify for accurate position of where the Spy device is within the environment.

\noindent\textbf{Joint Optimization.} To improve the accuracy of angle-of-arrival (AoA) estimation in a given spatial scenario, we exploit multiple channel measurements obtained under different dimensions, such as frequency, time, or deliberately introduced offsets. Suppose we have $L$ such measurements, each providing an IFFT-based spatial profile denoted as $P_i$. 

For each measurement, the conventional sparse formulation seeks an AoA profile $a_i$ via:
\[
\min_{a_i} \; \|P_i - W_i a_i\|_2^2 + \beta \|a_i\|_1,
\]
where $W_i$ is the dictionary of window functions of $i^{th}$ measurement. However, to enforce consistency across measurements---i.e., that they share the same underlying set of active angles---we adopt a joint optimization framework. We stack all sparse codes as columns of a matrix:
\[
A = [a_1, a_2, \ldots, a_L],
\]
and formulate the following group-sparse objective:
\[
\min_{A} \; \sum_{i=1}^{L} \|P_i - W_i a_i\|_2^2 + \lambda \sum_{j=1}^{M} \|A_{j,:}\|_2,
\]
where $M$ is the number of candidate angles. The mixed-norm regularization term $\sum_j \|A_{j,:}\|_2$ promotes \emph{row-wise group sparsity} in $A$, ensuring that entire rows are either all zeros or all non-zeros across samples. This induces a shared support set of active angles, thereby improving estimation robustness under diverse channel conditions. In addition, since those measurements are independent, the constraints of the joint optimization problem are simply the superposition of constraints of each sub-problem.

\noindent\textbf{Additional Benefits of Joint Optimization.}
Beyond improving AoA accuracy through shared-support recovery, our joint optimization framework also enables cross-dimensional analysis of the channel. As noted earlier, the non-zero entries in the recovered AoA profiles are proportional to
\[
    \frac{w_k}{\sum_{k=1}^{m} w_k}
\]
which correspond to the relative channel gain of each propagation path. Tracking how these non-zero coefficients vary across the $L$ measurements, therefore, provides additional insights into the propagation environment:
\begin{itemize}[leftmargin=*]
    \item \emph{Temporal diversity:} When measurements are collected at different time instants, variations in the coefficients at a given AoA reflect the temporal evolution of each propagation path, enabling the tracking of slow fading or device mobility.

    \item \emph{Frequency diversity:} When measurements span multiple frequencies, the coefficient variations encode the underlying path delays, since frequency-dependent phase shifts correspond to time-difference-of-arrival (TDOA) information.

    \item \emph{Offset diversity:} When deliberate time offsets are applied during measurement, differences in coefficient behavior help distinguish interfering signals that exhibit different autocorrelation structures.
\end{itemize}

Together, the joint framework not only enforces a consistent set of active angles across all measurements, but also provides a principled way to characterize how propagation paths evolve across time, frequency, and offset dimensions.

% Beyond improving the accuracy of AoA estimation through shared support recovery, our joint optimization framework offers additional benefits by enabling cross-dimensional analysis of the channel. As discussed earlier, the non-zero entries in the recovered AoA profiles are proportional to $\frac{w_k}{\sum_{k=1}^{m} w_k}$, effectively representing the relative channel gain along each estimated path.

% By examining the variation of these non-zero coefficients across the $L$ measurements, we can extract further insights about the channel:

% \begin{itemize}
%   \item \emph{Temporal diversity:} When measurements come from different time instants, the coefficient variations at a given AoA reveal the time evolution of each propagation path, enabling tracking of slow fading or mobility.
%   \item \emph{Frequency diversity:} When measurements are taken across different frequencies, the variations encode time-delay characteristics of the paths (TDOA), as frequency-dependent phase changes reflect path delays.
%   \item \emph{Offset diversity:} When measurements was offset differently, the differences in coefficient behavior can help distinguish interfering signals with different autocorrelation properties.
% \end{itemize}

% In essence, the joint framework not only enforces a consistent set of active angles, but also provides a structured way to analyze how channel characteristics evolve across diverse measurement dimensions.

%% file: evaluation/eva_morty.tex
\input{evaluation/implement_morty}

\input{evaluation/result_morty}

%% file: evaluation/implement_morty.tex
\section{Implementation and Setup}\label{sec:implementation}

\noindent\textbf{Hardware and Software.}
Our hardware platform consists of a switched antenna array controlled by a custom PCB for high-speed switching during emanation signal collection. The antenna array is built using commercial off-the-shelf (COTS) antennas and a PE42582 SP8T RF switch evaluation board~\cite{PE42582_datasheet}, which supports an operational frequency range of 9~kHz-8~GHz. This allows fast switching within the coherence time of the emanation signal, driven by external control logic that toggles among the antenna ports. The evaluation board provides eight RF input ports connected to the antennas and one RF common port connected to a USRP~B210 receive chain. All antennas are mounted on laser-cut acrylic plates supporting up to eight elements in any configuration. The calibration trace on the evaluation board routes one end to the reference antenna and the other to the B210’s second receive chain, enabling simultaneous reference IQ capture for signal identification.

We design a custom PCB to control the RF switch and supply the required bias voltages. On-board LDOs generate a 3.4V supply (V\textsubscript{DD}) and a -3V bias voltage for rapid switching. A CMOD~A7-35T FPGA handles antenna-port selection. The FPGA implements a finite-state machine (FSM) clocked at 40~MHz and synchronized using trigger and reset signals from the USRP~B210.

The control logic ensures that none of the eight input ports is connected to the common port while the system is idle. When the reset signal is logic low, the FSM waits for a trigger pulse to begin the switching sequence. Upon receiving the trigger, the FSM cycles through antennas by asserting a 3-bit control signal (V1-V3) corresponding to the selected input port. The switching period is determined by a configurable parameter, \emph{switch\_time}, which is sent from a Python interface to the FPGA via UART before each measurement run. During each \emph{switch\_time} interval, the 3-bit control signal increments by one, selecting the next antenna in sequence. When the reset line is asserted high, the FSM halts and returns to its initial state. Our signal processing algorithms are mainly implemented in Python. 

\begin{figure}
    \centering
    \includegraphics[width=\linewidth]{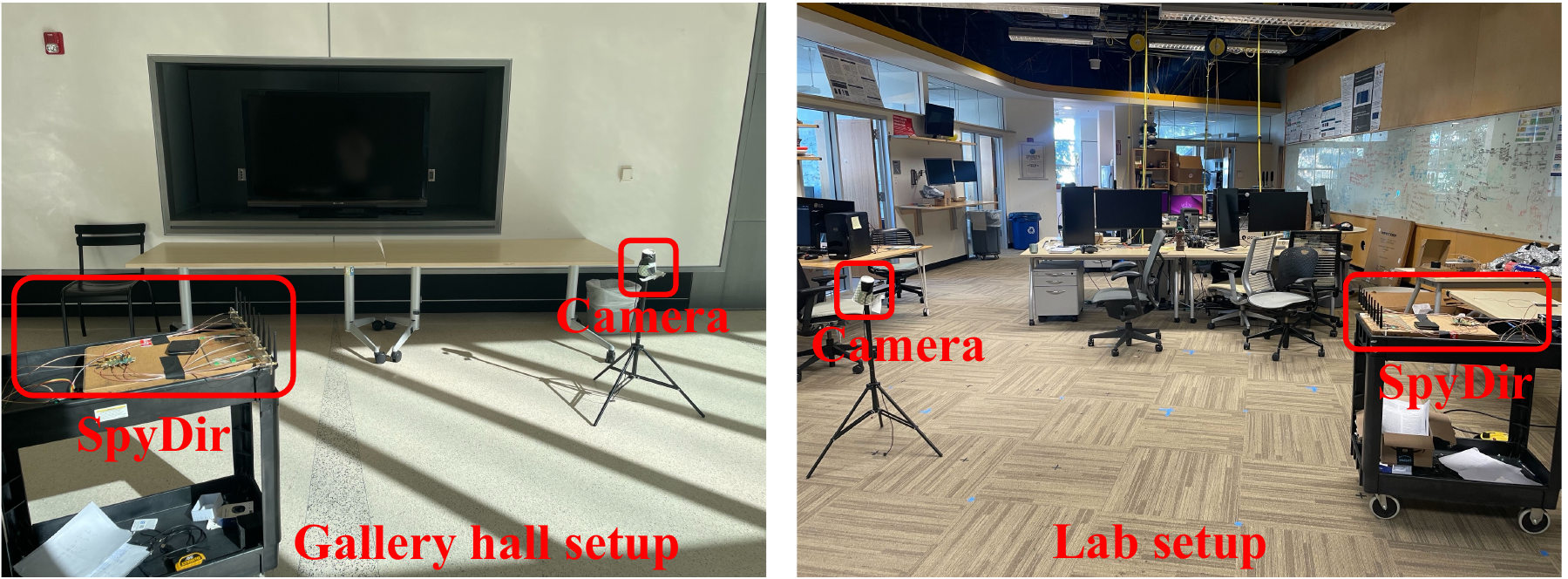}
    \caption{Real-world experimental evaluation of \name’s switched-antenna receiver for hidden spy camera AoA estimation in the gallery hall and lab.}
    \label{fig:res-deployment}
\end{figure}

\noindent \textbf{Experimental settings.}  
Both USRP RX ports operate at a 3.072MS/s sample rate with a receive gain of 75 dB. A uniform linear array (ULA) of eight switched antennas is used, with an inter-element spacing of 62.5mm to balance array size and frequency-dependent mismatch. The RF switch cycles through all eight antennas within a total switching interval of \(31.25~\text{ms} \times 8\). A Linux machine equipped with an Intel i7-1185G7 CPU running Ubuntu~20.04 drives the USRP and performs subsequent signal processing using Python-based implementations.

\begin{figure*}[t]
\centering
\includegraphics[width=0.9\linewidth]{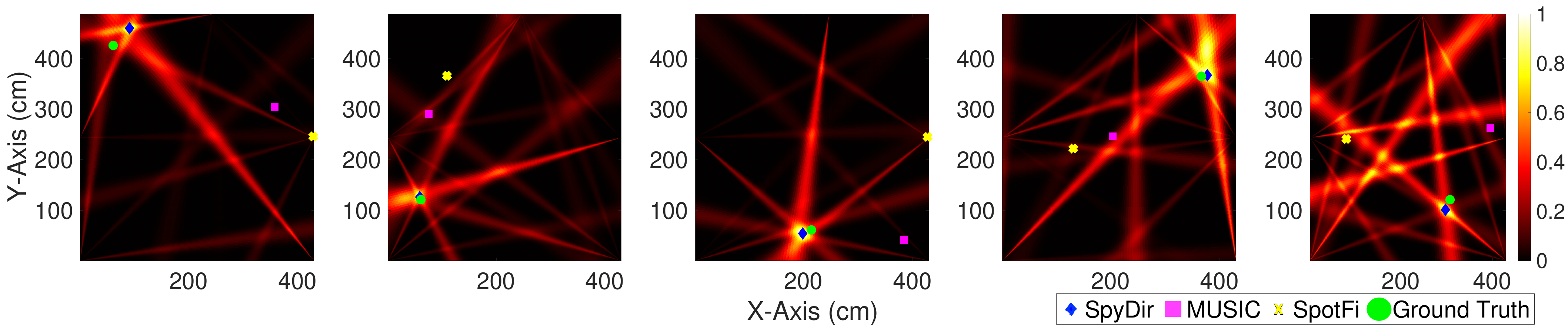}
\vspace{-1em}
    \caption{Ground Truth and the Triangulated 2D location results using \name of the Spy camera Devices when placed in a real-world lab setup where there are multiple interferences at locations 1-5 (from left to right).}
    \label{fig:res-triangulation}
\end{figure*}

We deploy \name in indoor environments with strong multipath. The hidden cameras are placed at distances between 2-4m from the \name setup. For each measurement run, we collect 50s of IQ samples from the emanations unintentionally emitted by the spy devices. From these samples, we identify emanation harmonics spread over the 1GHz sub-GHz band, which are then used for 
AoA estimation.

We evaluate \name using two representative commercial spy-camera devices (OV2640~\cite{esp32-cam} and HZVS4730~\cite{esp32-timer}) that do not actively transmit RF signals but nonetheless emit continuous electromagnetic leakage. Experiments are conducted in multipath-rich and non-line-of-sight (NLoS) scenarios, including deployment behind walls, as illustrated in Fig.~\ref{fig:res-deployment}. We compare our AoA estimation performance against state-of-the-art methods, including SpotFi~\cite{kotaru2015spotfi} and the MUSIC algorithm~\cite{friedlander1990sensitivity}.

%We have demonstrated that \name can provide an average AoA estimation error of \todo{$z^\circ$} and a maximum AoA estimation error of \todo{$zz^\circ$}. \name demonstrates \todo{$x\%$} better results compared to using without averaging, and \todo{$y\%$} better results compared to using SpotFi instead of \name's optimization algorithm.

%For the clock leakage signal, we focus on one single spike in the 1G-2G range where interference is minimal, using the extracted narrow-band signal for AOA localization. 

%% file: evaluation/result_morty.tex
\section{Evaluation}\label{sec:results}

We evaluate \name across three dimensions:
(1) a real-world 2D localization case study comparing triangulated results with baselines;
(2) end-to-end AoA accuracy benchmarking across algorithms, device types, and averaging strategies;
(3) targeted microbenchmarks evaluating the effects of range, time-offset noise decorrelation, time-averaging window size, and sparsity parameter $\beta$ on AoA performance. We mainly report the localization error, AoA estimation error, and the emanation SNR. All experiments use the hardware and deployment setup described in~\S\ref{sec:implementation}.

\subsection{Case Study: 2D Localization}

% \begin{figure*}[t]
%     \begin{minipage}{0.25\linewidth}
%         \includegraphics[width=\linewidth]{figures/plot_sensys/loc_error_s2.pdf}
%         \subcaption{Location-1}
%     \end{minipage}
%     \begin{minipage}{0.25\linewidth}
%         \includegraphics[width=\linewidth]{figures/plot_sensys/loc_error_s3.pdf}
%         \subcaption{Location-2}
%     \end{minipage}
%     \begin{minipage}{0.25\linewidth}
%         \includegraphics[width=\linewidth]{figures/plot_sensys/loc_error_s4.pdf}
%         \subcaption{Location-3}
%     \end{minipage}
%     \begin{minipage}{0.25\linewidth}
%         \includegraphics[width=\linewidth]{figures/plot_sensys/loc_error_s5.pdf}
%         \subcaption{Location-4}
%     \end{minipage}
%     \begin{minipage}{0.25\linewidth}
%         \includegraphics[width=\linewidth]{figures/plot_sensys/loc_error_s6.pdf}
%         \subcaption{Location-5}
%     \end{minipage}
%     \caption{Ground Truth and the Triangulated 2D location results using \name of the Spy camera Devices when placed in a real-world lab setup where there are multiple interferences at locations 1-5.}
%     \label{fig:res-triangulation}
% \end{figure*}

We demonstrate how \name can be deployed in a real-world indoor environment to locate a hidden spy camera.

\noindent\textbf{Method:}
We place the spy camera at five different locations within a $5\,\text{m} \times 5\,\text{m}$ lab space containing numerous electronic devices and IoT equipment, all of which serve as strong sources of interference. \name identifies the target device by leveraging its unique clock frequency and the self-correlation–based interference detection mechanism described in~\cref{subsec:interference}. 

To obtain multiple vantage points for triangulation, we position \name’s switched-antenna receiver at five different locations throughout the lab. At each vantage point, we collect leakage measurements and estimate the AoA of the spy camera. The final 2D position estimate is obtained by triangulating these AoA results.

\noindent\textbf{Results:}
Fig.~\ref{fig:res-triangulation}(a–e) shows the triangulation performance across five target locations. 
Table~\ref{tab:localization} reports the localization error for the five target locations. \name achieves an average error of 19.86 cm, with individual errors ranging from 5.39 cm to 46.01 cm. In four of the five cases, the error remains within 5–25 cm, demonstrating stable high accuracy even under strong interference from surrounding IoT devices. The localization error of 46.01 cm is mainly because the spy camera is at the edge of the aperture range of the switched-antenna array, resulting in noisy emanations. 

In contrast, MUSIC and SpotFi perform substantially worse, with average errors of 206.79 cm and 294.75 cm, respectively. Aggregating across all deployments, \name improves localization accuracy by 10.4× over MUSIC and 14.8× over SpotFi, highlighting that classical AoA algorithms cannot reliably operate on weak, unintentional emanations, whereas \name’s sparse optimization and noise-decorrelation pipeline successfully recovers fine-grained directional information.

\begin{figure*}[t]
     \centering
     \begin{minipage}{0.3\linewidth}
         \includegraphics[width=\linewidth]{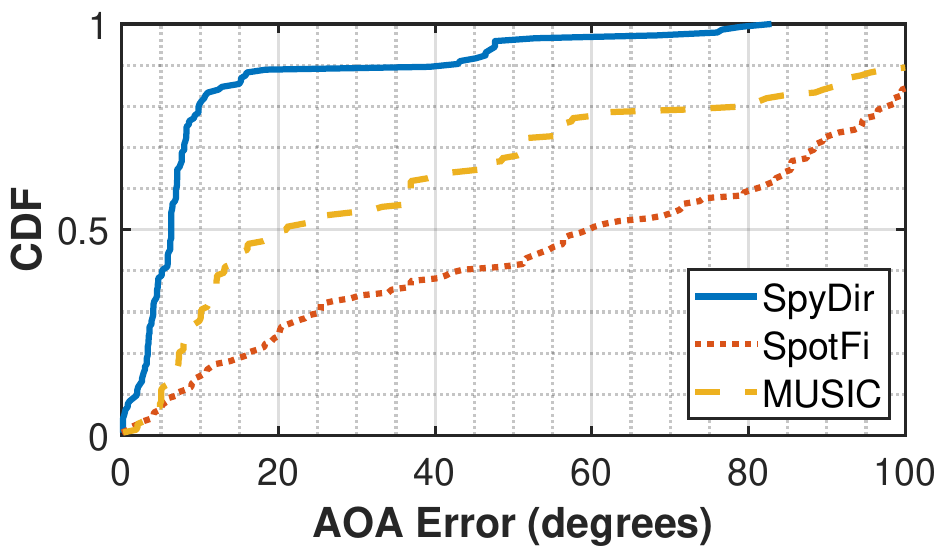}
         \subcaption{Algorithm comparison.}
     \end{minipage}
     \begin{minipage}{0.3\linewidth}
         \includegraphics[width=\linewidth]{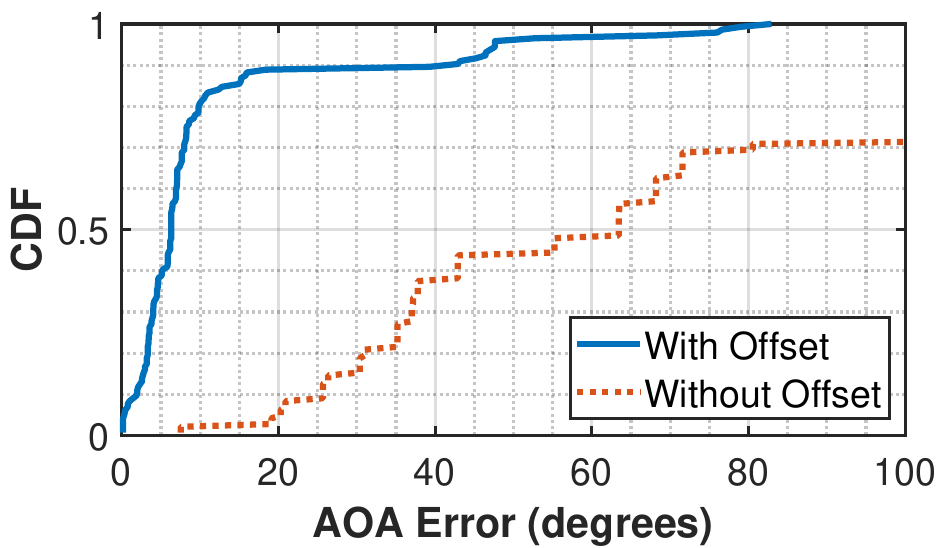}
         \subcaption{Effect of time offset.}
     \end{minipage}
     \begin{minipage}{0.3\linewidth}
         \includegraphics[width=\linewidth]{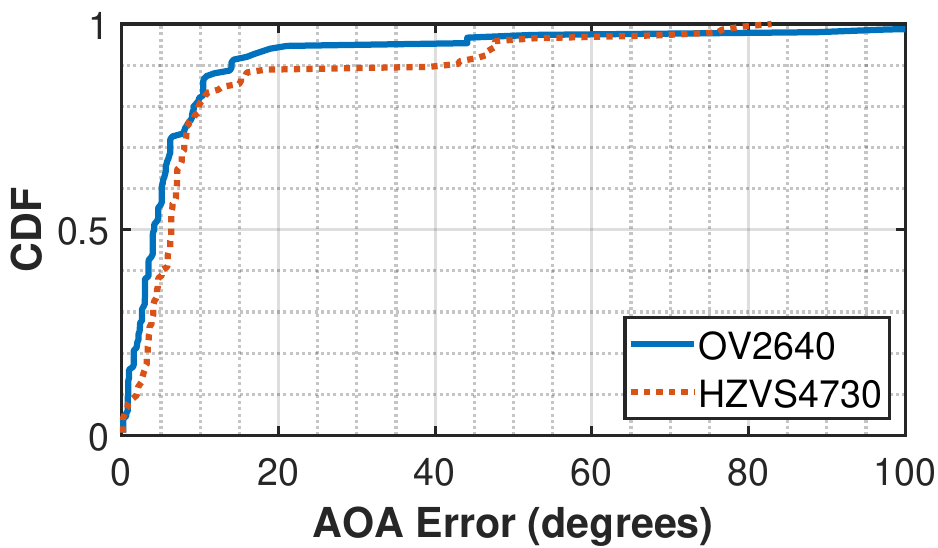}
         \subcaption{Across different devices.}
     \end{minipage}
     \caption{CDF of AoA estimation error for (a) algorithm comparison, 
              (b) time-offset–based noise decorrelation, 
              and (c) heterogeneous spy-camera devices.}
     \label{fig:AOA_results}
\end{figure*}

\begin{table}[t]
\centering
\caption{Localization error across five target locations.}
\label{tab:localization}
\begin{tabular}{lccc}
\toprule
\textbf{Location} & \textbf{\name\ (cm)} & \textbf{MUSIC (cm)} & \textbf{SpotFi (cm)} \\
\midrule
1 & \textbf{46.01} & 320.16 & 408.77 \\
2 & \textbf{5.39}  & 170.58 & 249.66 \\
3 & \textbf{17.46} & 171.17 & 281.47 \\
4 & \textbf{11.18} & 201.82 & 275.09 \\
5 & \textbf{22.36} & 166.21 & 256.77 \\
\midrule
\textbf{Avg.} & \textbf{19.86} & 206.79 & 294.75 \\
\bottomrule
\end{tabular}
\end{table}

\subsection{End-to-End AoA Performance}\label{sec:res-devices}

To comprehensively evaluate the AoA estimation performance of our proposed system, we conduct extensive experiments under diverse multipath-rich indoor environments. Our evaluation focuses on three key aspects: 
\begin{itemize}[leftmargin=*]
\item We compare the AoA estimation accuracy across different algorithms, including traditional MUSIC, SpotFi, and our proposed approach. 

\item We investigate the impact of time-offset–based noise decorrelation technique on system performance. 

\item We examine the generalization capability across heterogeneous spy-camera devices to demonstrate the robustness and practicality of our system.
\end{itemize}

\noindent \textbf{Result:} 
Fig.~\ref{fig:AOA_results}(a) presents the CDF of the AoA error for \name, SpotFi, and MUSIC. \name achieves an average error of only \ang{6.30}, substantially outperforming SpotFi (\ang{59.89}) and MUSIC (\ang{21.06}). This improvement stems from \name's sparse matrix optimization–based AoA solver, which remains stable even when the channel matrix is low rank—a regime where both SpotFi and MUSIC degrade sharply.

Fig.~\ref{fig:AOA_results}(b) evaluates \name with and without applying the time-offset technique. Introducing the offset reduces the average error from \ang{63.43} to \ang{6.30}, demonstrating that it effectively decorrelates noise across switching-antenna ports and enables \name to extract fine-grained AoA cues from extremely weak emanation signals.

Fig.~\ref{fig:AOA_results}(c) reports AoA accuracy across different spy-camera devices. The error remains consistently low regardless of device model, indicating that \name is hardware-agnostic and generalizes across heterogeneous IoT camera platforms. Because \name relies only on the emanation structure rather than device-specific features, it scales naturally to diverse real-world spy camera deployments.

\noindent\textbf{Remark:} The above results demonstrate that \name reliably delivers accurate AoA estimation for spy-camera emanations even in multipath-rich indoor environments. This performance arises from three key technical contributions: (1) sparse matrix optimization for robust AoA derivation, (2) non-coherent temporal averaging to enhance weak emanations, and (3) time-offset–based noise decorrelation across switching antennas. Importantly, \name requires only the periodicity of the emanation signal and imposes no assumptions on its waveform, enabling broad applicability for hidden-camera detection. We believe this work advances emanation-based source localization and represents a step toward practical, privacy-preserving defense against covert surveillance.
\looseness-5

\subsection{Microbenchmarks}

\begin{figure*}
\centering
\begin{minipage}{0.24\textwidth}
  \centering
      \captionsetup{width=0.95\textwidth}

    \includegraphics[width=\linewidth]{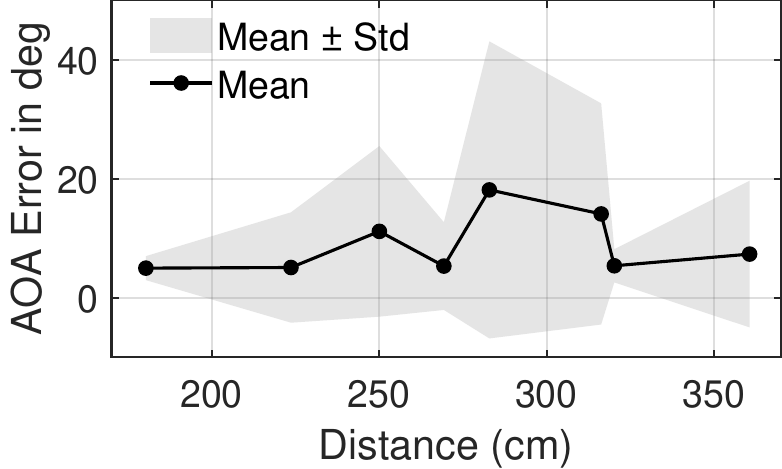}
    \caption{AoA estimation error over different distances between the spy device and \name.}
    \label{fig:res-range-aoa}
\end{minipage}%
\begin{minipage}{0.24\textwidth}
  \centering
      \captionsetup{width=0.95\textwidth}

    \includegraphics[width=\linewidth]{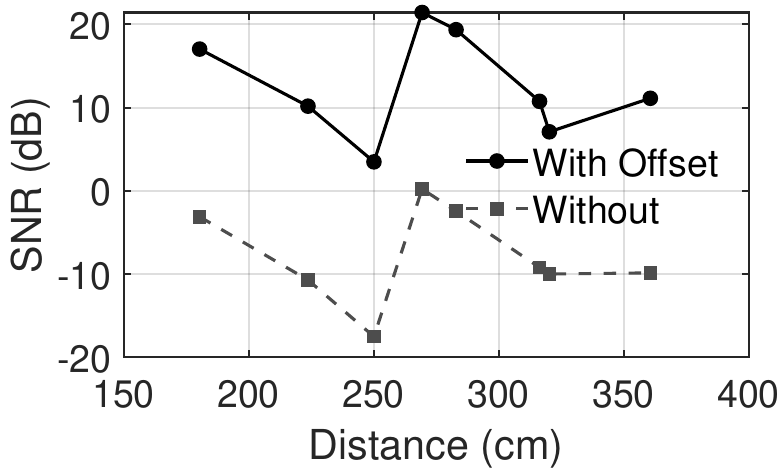}
    \caption{SNR of the sniffed emanations over different distances with and without time offset. }%, and (b) frequency bands of interest averaging for AoA estimation.}
    
    \label{fig:res-range}
\end{minipage}%
\begin{minipage}{0.24\textwidth}
  \centering
      \captionsetup{width=0.95\textwidth}

     \includegraphics[width=\linewidth]{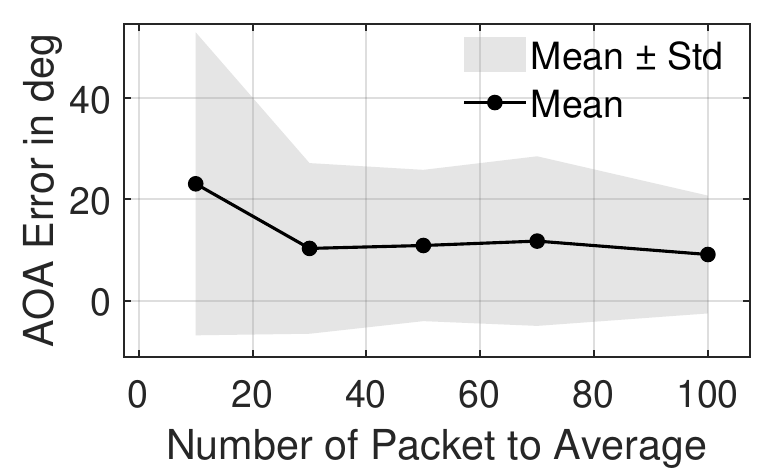}
    \caption{AoA estimation error over different numbers of averaging packets. Each packet last for 50ms.}
    \label{fig:res-window}
\end{minipage}%
\begin{minipage}{0.24\textwidth}
  \centering
      \captionsetup{width=0.95\textwidth}
      
    \includegraphics[width=\linewidth]{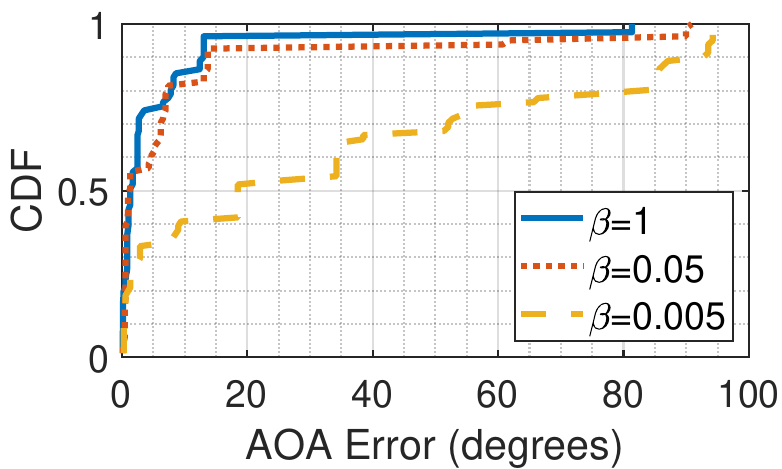}
    \caption{CDF of AoA estimation error for various value of $\beta$ (super parameter in Equation~\ref{eq:optimization}).}
    \label{fig:impact-beta}
\end{minipage}
\end{figure*}

\subsubsection{Impact of Range}
The distance between the spy camera and \name affects the received leakage strength and, consequently, the AoA estimation accuracy.

\noindent\textbf{Method:}
We place the spy camera at varying distances from \name in a real indoor environment and measure the resulting AoA estimation error.

\noindent\textbf{Result:}
Fig.~\ref{fig:res-range-aoa} reports the mean and standard deviation of AoA error across distances. 
AoA accuracy remains stable up to approximately 200\, cm, beyond which the error increases sharply (e.g., at 225\, cm), revealing \name's practical operating range. 
This degradation stems from reduced leakage strength and a stronger influence of multipath at low SNRs. 
Within the 200\, cm range, however, \name remains robust and maintains consistently low AoA error, since we enhance the emanation SNR through time-offset compensation.

\subsubsection{Impact of Time Offset}
\name uses non-coherent averaging across packets, which benefits from noise decorrelation achieved via our time-offset mechanism.

\noindent\textbf{Method:}
We compute the relative SNR (emanation spike amplitude over noise floor) at various distances, comparing measurements with and without the time offset.

\noindent\textbf{Result.}
As shown in Fig.~\ref{fig:res-range}, the time offset provides a dramatic and consistent SNR improvement across all distances. Without the offset, SNR falls from ~10\,dB to below --10\,dB, making the leakage nearly undetectable. With the offset, SNR remains above 20\,dB at short distances and above 10\,dB at the farthest point, offering a \textbf{10--25\,dB gain}.  
This large improvement arises from noise decorrelation across the switched antenna ports, enabling non-coherent averaging to effectively enhance the periodic leakage while suppressing noise. Consequently, \name maintains usable SNR for AoA estimation at significantly longer ranges.

\subsubsection{Impact of Time-Averaging Window Length}
Because leakage signals are extremely weak, increasing the averaging window substantially improves AoA accuracy. We employ non-coherent averaging to strengthen the periodic leakage component while suppressing unstructured noise and evaluate its effect on AoA performance.

\noindent\textbf{Method:}
We vary the number of averaged packets (each packet is 50\,ms) to study how strengthening the leakage signal affects AoA accuracy. For each window size, we average the IQ samples and extract the leakage spikes. The structured, periodic nature of clock leakage persists across packets, whereas noise averages out, making non-coherent averaging effective.

\noindent\textbf{Result.}
Fig.~\ref{fig:res-window} shows that both the mean and standard deviation of AoA error decrease as more packets are averaged.  
Averaging 100 packets reduces the AoA error by \ang{13.9} and lowers the standard deviation by 60\% compared to averaging 10 packets.  
This improvement arises because the periodic leakage reinforces across packets, whereas unstructured noise is suppressed by averaging. The stronger reconstructed leakage then leads to significantly more accurate AoA estimation.

\subsubsection{Impact of $\beta$}
During the experiment, we observed that \name's AoA estimation relies on a sparsity-regularized optimization (Eq.~\ref{eq:optimization}), where the hyperparameter $\beta$ controls the sparsity level.

\noindent\textbf{Method.}
Using controlled simulations of multipath-rich indoor environments, we vary $\beta$ and measure the resulting AoA estimation error. 
We then apply the empirically optimal value $\beta=1$ in real experiments.

\noindent\textbf{Result.}
Fig.~\ref{fig:impact-beta} shows the CDF of AoA estimation error for different values of $\beta$. The choice of $\beta$ directly controls the sparsity of the recovered AoA profile, and thus overly small or overly large values degrade accuracy. When $\beta$ is too small (e.g., $\beta = 0.005$), the solution becomes insufficiently regularized, allowing noise to introduce many spurious AoA components and leading to large errors. Conversely, when $\beta$ is too large, the optimization becomes over-sparse and suppresses valid paths, causing the estimated channel to collapse toward zero.

Across the tested values, $\beta = 1$ provides the best balance between sparsity and fidelity, resulting in the lowest AoA estimation error. Therefore, \name adopts $\beta = 1$ as the empirically optimal value for all deployments.

%% file: related/related_morty.tex
\section{Related Work}

\textbf{Dedicated detectors.}
A straightforward way to detect hidden IoT devices is to use dedicated detectors~\cite{ded:detector, anti_detector, anti_detector1, ded:detector2}, which require users to manually sweep the detector around the environment. These devices sense harmonics generated by electronic components through near-field coupling between the detector and the device’s circuitry. As a result, they impose substantial human effort and cannot support automated or fine-grained localization. In contrast, \name performs automated detection and localization using emanations passively emitted by powered-on IoT devices, without requiring any active transmission or handheld scanning.

\noindent \textbf{Camera and laser sensor-based detection.}
Hidden cameras pose a significant privacy threat, and many prior approaches detect them using optical sensors such as cameras or lasers~\cite{he2018active, li2015fast, qian2015recognition, sadler2010mobile, svedbrand2019optics, zhang2017fast, liu2019spectrum, liu2019analysis, liu2019design, sami2021lapd}. These methods exploit visible hardware features such as lens reflections or LED indicators to identify or localize cameras. However, such approaches depend on line-of-sight visibility and on the presence of detectable optical components, making them ineffective when devices are concealed behind objects, embedded in furniture, or otherwise visually obscured.

\noindent\textbf{Traffic-based detection.}
Hidden IoT devices equipped with radio transceivers often generate wireless traffic, enabling a variety of traffic-analysis-based detection approaches~\cite{cheng2018dewicam,singh2021always,he2021motioncompass, liu2018detecting,salman2022csi,miettinen2017iot, cheng2019detecting, mitev2020leakypick,li2018adversarial, lagesse2018automated, wu2019you, sharma2022lumos}. For example, MotionCompass~\cite{he2021motioncompass} detects hidden cameras by correlating their transmitted traffic with a user’s motion trajectory. However, such approaches provide only coarse-grained localization and require carefully planned user movement. In contrast, \name does not rely on wireless traffic at all; it operates solely on unintentional electromagnetic emanations and can detect and localize spy devices even when they are not actively transmitting.

\noindent\textbf{Emanation-based detection.}
Recent work leverages unintentional electromagnetic emanations to detect hidden IoT devices in indoor environments, since such emanations are produced automatically by virtually all electronic devices~\cite{shen2021earfisher,liu2023camradar, sun2025revealing, li2018eye, zhou2023dehirec}. For example, RFScan~\cite{sun2025revealing} detects hidden devices by scanning for their emanation harmonics and attempts localization using a directional antenna. However, its accuracy degrades severely indoors because the method depends heavily on received signal strength and is highly susceptible to multipath effects. Similarly, ESauron~\cite{zhang2024eye} relies on the RSSI of electromagnetic radiation for detection but faces analogous challenges under complex propagation conditions. Although CamFirm~\cite{liao2025camfirm} advances the state of the art by actively modulating ambient brightness to elicit camera-specific radiation for localization, its operation is confined to dim environments and does not generalize to typical indoor settings. More critically, all these approaches follow a search-based paradigm—requiring physical movement within the environment and offering only coarse-grained location estimates—which fundamentally restricts their real-world utility. More broadly, most emanation-based systems focus solely on detection and do not provide precise localization of the hidden cameras. In contrast, SpyDir addresses this gap by using a custom switched-antenna array and a novel optimization-based AoA estimator that enables accurate localization directly from emanations

\noindent\textbf{AoA estimation.}
AoA estimation has been extensively studied in indoor wireless localization, with techniques through antenna array ranging from geometry-based methods~\cite{chuang2015high} and FFT-based estimators~\cite{schwarz1990use} to subspace methods such as MUSIC~\cite{friedlander1990sensitivity} and hybrid approaches like SpotFi~\cite{kotaru2015spotfi}. However, these methods assume access to actively transmitted signals with known pilots or high-quality CSI, conditions that do not hold for unintentional emanations. In \name's task, the emanations are extremely weak and produce a low-rank relative channel matrix, making traditional AoA techniques ineffective. To overcome these limitations, \name introduces a sparse optimization–based AoA estimator specifically designed for weak, pilotless emanations captured via a switched-antenna array.

%% file: discussion.tex
\vspace{-2em}
\section{Discussion}

\noindent\textbf{Practical Insights and System Behavior.}
\name focuses on the problem of localizing hidden spy devices using their unintentional electromagnetic emanations. While \name benefits from existing emanation-detection techniques, the design of robust detection algorithms is outside the scope of this work. Prior efforts~\cite{sun2025revealing, li2018eye, zhou2023dehirec} provide effective methods for identifying weak leakage signatures, and \name can operate in conjunction with any such detector.

A key contribution of \name is its ability to extend the practical detection range up to 200cm, a 120\% improvement over state-of-the-art approaches. More importantly, \name achieves accurate direction finding from weak, pilotless emanations, with a median AoA error of $4.87^\circ$ and a 90th-percentile error of $15.08^\circ$. This demonstrates that accurate localization of devices emitting extremely low-power leakage is feasible even in multipath-rich indoor environments.

\noindent\textbf{Limitations and Future Work}
Although \name tolerates interference from other IoT devices, it still depends on the underlying detector to identify the correct emanation frequency of a target device. Existing detectors focus mainly on clock-based leakage from hidden cameras, and extending this capability to a broader set of IoT devices (e.g., microphones with significantly weaker leakage) remains an open challenge. A systematic characterization of leakage frequencies across different commercial devices, especially the weaker leakage from microphones, would further improve the robustness of emanation-based localization.

In addition, the wideband nature of sub-GHz leakage requires a relatively large antenna aperture to achieve sufficient angular resolution. This introduces a physical footprint that is limited by the capabilities of commodity SDR hardware. Future work will explore (i) antenna-placement strategies that emulate larger apertures in compact form factors~\cite{zheng2019extended}, and (ii) detection and utilization of higher-frequency leakage components, which would naturally support smaller array geometries.